\documentclass[conference]{IEEEtran}
\IEEEoverridecommandlockouts
\usepackage{cite}
\usepackage{amsmath,amssymb,amsfonts}
\usepackage{algorithmic}
\usepackage{graphicx}
\usepackage{textcomp}
\usepackage{xcolor}
\usepackage{balance} 
\usepackage{enumitem}
\usepackage{booktabs}
 \usepackage{subcaption}
\usepackage{comment}
\def\BibTeX{{\rm B\kern-.05em{\sc i\kern-.025em b}\kern-.08em
    T\kern-.1667em\lower.7ex\hbox{E}\kern-.125emX}}
\newcommand{\unnumberedparagraph}[1]{\paragraph*{#1}\mbox{}} 

\begin{document}
\title{A Large-Scale Feasibility and Ethnography Study of Screen-based AR and 3D Visualization Tools for Anatomy Education: Exploring Gender Perspectives in Learning Experience}

\author{\IEEEauthorblockN{Roghayeh Leila Barmaki$^{1}$$^{\S}$\thanks{$^{\S}$ The study was conducted during the postdoctoral fellowship of Roghayeh Leila Barmaki at Johns Hopkins University. Correspondence to rlb@udel.edu.}, 
Kangsoo Kim$^{2}$,  
Zhang Guo$^{1}$, 
Qile Wang$^{1}$, \\
Kevin Yu$^{3}$,
Rebecca Pearlman$^{4}$, and
Nassir Navab$^{3, 4}$ 
}

\IEEEauthorblockA{$^{1}$ University of Delaware, Newark, DE, USA
}
\IEEEauthorblockA{$^{2}$ University of Calgary, Calgary, AB, Canada}

\IEEEauthorblockA{$^{3}$ Technical University of Munich, Munich, Germany	
}
\IEEEauthorblockA{$^{4}$Johns Hopkins University, Baltimore, MD, USA
}
}

\maketitle

\begin{abstract}
    
Anatomy education is an indispensable part of medical training, but traditional methods face challenges like limited resources for dissection in large classes and difficulties understanding 2D anatomy in textbooks. Advanced technologies, such as 3D visualization and augmented reality (AR), are transforming anatomy learning. This paper presents two in-house solutions that use handheld tablets or screen-based AR to visualize 3D anatomy models with informative labels and in-situ visualizations of the muscle anatomy. To assess these tools, a user study of muscle anatomy education involved 236 premedical students in dyadic teams, with results showing that the tablet-based 3D visualization and screen-based AR tools led to significantly higher learning experience scores than traditional textbook. While knowledge retention didn't differ significantly, ethnographic and gender analysis showed that male students generally reported more positive learning experiences than female students. This study discusses the implications for anatomy and medical education, highlighting the potential of these innovative learning tools considering gender and team dynamics in body painting anatomy learning interventions.
\end{abstract}
\section{Introduction} \label{Sec:Intro}
\noindent
Human anatomy and physiology are vital parts of medical education that involve complex functional structures and movements of the human body. 
Comprehensive learning of anatomy and physiology provides a thorough understanding of human body function, enabling more effective treatment of abnormal or disease states~\cite{BLANCHARD200573}.
However, the complexity of the course poses challenges for students in achieving their desired learning outcomes.
Several factors associated with the learning experience can influence these outcomes, including the learning tools, the quality of the material, the student's prior experience, and their emotional concerns~\cite{o2008development, green2018relationship, chan2019approaches}.


\noindent The most common practice for students in anatomy education is to use textbooks with static images, but this cannot provide the students with a realistic first-hand and interactive experience, which may not be effective for their learning experience and performance~\cite{leung2020modernising}.
Despite the effectiveness of traditional methods for training anatomy, such as dissection or prosection, these methods have become less feasible nowadays due to limited resources, large class sizes, and mainly the absence of face-to-face learning experiences, to name a few constraints.
Due to such limitations, most medical, dental, and other allied health schools have recently declined the practical laboratory hours for anatomy\cite{leung_anatomy_2006,winkelmann_anatomical_2007}.
In anatomy and physiology education, spatial visualization is likely essential for students to learn the dynamics of anatomical structures and spatial relationships to surrounding structures.
The traditional visualization of human anatomy in 2D textbook views insufficiently captures the complexity of human anatomy as students are often required to mentally reconstruct 3D spatial relationships, which presents a considerable challenge.

Virtual/augmented reality (VR/AR) can provide information on dynamics and spatial relationships interactively and intuitively by employing 3D virtual skeletons and organs and adding a virtual information layer on top of the physical body.
While various medical training scenarios have used these technologies~\cite{lovis_mixed_2020,marmulla_augmented_2005,Romand2020}, the use of computer-generated 3D models, in particular, allows students to rotate and locate structures from various views and perspectives in anatomy learning. 
Such dynamic visualization techniques improve student visual-spatial abilities\cite{huk_who_2006,lipponen_challenges_1999,stieff_mental_2007}.
Moreover, virtual 3D visualizations offer more accessible opportunities to engage and explore anatomical structures than traditional cadaver-based learning for their repeatability and monitoring capabilities.


Body painting has been shown as an effective tool for learning anatomy and associated clinical skills \cite{cookson2018exploration,Diaz2021learning}. It is a motivating and creative experience for students that provides memorable visual images and encourages multisensory and active participation. 
While body painting suits all students, cultural sensitivity, gendered considerations, and careful negotiation may be necessary to ensure all students are comfortable carrying out the activities.

Inspired by renown methods of anatomy education, in this work, we propose a user study in an actual educational laboratory setting that uses 3D visualizations in an anatomy body painting learning task.
In a controlled, team-based large-scale study with $236$ participants, we compare our in-house tablet-based 3D visualizations (\emph{Tablet-3D}), and our screen-based AR (\emph{Screen-AR})~\cite{barmaki_enhancement_2019} with the conventional paper-based \emph{Textbook}. We aimed to find any interplay between the participants' performance outcomes with learning tools, gender, and group gender compositions. 
Our research aims to address the following research questions:
\begin{itemize}[leftmargin=*]
    \item \textbf{RQ1:} Do tablet-based visualizations and AR technology improve students' learning experience compared to the traditional textbook in anatomy education?
    \item \textbf{RQ2:} Do tablet-based visualizations and AR technology increase the learning outcomes, e.g., test scores or knowledge retention, in anatomy education?
    \item \textbf{RQ3:} Are there any particular benefits of AR technology in anatomy education experience over tablet-based visualizations?
    \item \textbf{RQ4:} Is the student's gender a factor to interplay with the effects of digital technology, or in general for the learning experience?
\end{itemize}
To answer these research questions, we first introduce our technological tools, tablet-based interactive visualization application and large screen-based AR tool, which can show dynamic anatomical information with interactive life-size 3D virtual models on top of a physical body.
We then report the findings from our large-scale study with $236$ students in teams of two who were participated in the body painting activity using three learning tools.

We found that students who used our Tablet-3D and Screen-AR conditions had more positive (anatomy) learning experiences than those who used a textbook, according to their self-reported outcomes.
In addition, we analyzed the potential effects of the participant's gender on the performance, which will be elaborated more in the following.

\section{Related Work}\label{Sec:RelatedWork}

\subsection{3D Technologies for Anatomy Education}


\noindent Anatomy is a complex subject that cannot be learned only from textbooks\cite{nainggolan2020user}.
Traditional anatomy training is based on the dissection and pro-section of the human body, which provides tangible haptic interactions and realistic environment settings\cite{snelling_attitudes_2003,gunderman_exploring_2005}.
Despite the advantages of dissection, it equally raises concerns.
Studies have shown that the learning outcomes and the quality of dissections may be affected by the quality of the material, students' prior experience, and emotional concerns\cite{trelease_going_2000,winkelmann_anatomical_2007}.
In particular, both inexperienced and experienced medical and healthcare students are frequently appalled by the fear of death and the unnatural smell of cadavers during the dissection\cite{mclachlan_teaching_2004,winkelmann_anatomical_2007}. 
Instead of using deceased bodies, clay models provide an alternative solution for educators.
DeHoff et al.\cite{dehoff_learning_2011} found that compared with animal dissections, students had a better learning experience with clay models.
However, clay models cannot present complex anatomical regions or the functional movements of the structures in the anatomical domain.
Additionally, they impose challenges in transportation and storage. 
Moreover, anatomy course lab hours have gradually decreased in the past decades \cite{leung_anatomy_2006,winkelmann_anatomical_2007}, bringing more challenges to anatomy education.

\noindent As a response to the barriers and changes over time, anatomical learning platforms increasingly adapt from traditional methods to digital technology. 
Like any other reshaping processes, some anatomy scholars argue that dynamic visualization compensates for students' low spatial abilities by providing an explicit external representation of the system\cite{hays_spatial_1996,huk_who_2006,stieff_mental_2007,chickness2022novel}.
Increasingly powerful and accessible computer hardware allow 3D visualizations to replace or supplement traditional teaching in healthcare regarding lectures, cadavers, and textbooks\cite{yammine_meta-analysis_2015,golenhofen_use_2019,lemos_design_2019,maresky_virtual_2019}.
Donnelly et al.\cite{donnelly2009virtual} investigated the use of Virtual Human Dissector© (VHD) software, interactive teaching tools for cross-sectional anatomy, capable of reconstructing 3D views from 2D images, in anatomy education with self-directed learning and found no significant difference when compared with a students group that learns from using prosection, models, and textbooks. 
Kennan \& Awadh \cite{keenan2019integrating} discussed the effective utilization of visual 3D learning technologies as self-learning resources in the context of cross-sectional anatomy.
They proposed integrating the use of the 3D VHD system with Sectra\cite{barrack2015step}, a medical imaging device, to enhance the understanding of cross-sectional anatomy.

Lim et al.\cite{lim_use_2016} used 3D-printed models instead of traditional cadaveric specimens during the learning of external cardiac anatomy. 
Although 3D printing technology can provide teaching materials, the 3D printed model is limited by the complexity of anatomical regions. 
Equally, researchers found that 3D visualization methods improved student performance by providing multiple anatomical views and different perspectives of 3D rotating models, even on 2D screens \cite{yammine_meta-analysis_2015}.
Mobile-based applications and web-based 3D games have also been used as efficient learning tools for the study of human skeletal, muscular, and cardiovascular systems to explore more spatial information about the 3D anatomical models \cite{golenhofen_use_2019,lemos_design_2019}.

VR and AR techniques have been adopted into anatomy education in recent years\cite{maresky_virtual_2019,bork_empirical_2017,silva_emerging_2018,bacca2014augmented}.
As dynamic visualizing tools, they engage students in an immersive environment with audio and visual interactions and stereoscopic 3D models to have a better functional understanding of the anatomical structure and its movement within the 3D body space\cite{hackett_effect_2018,jacob2012lindsay,luursema_role_2008,preim2018survey}.
Duncan‐Vaidya and Stevenson\cite{duncan-vaidya_effectiveness_2020}
have found that experience from AR positively influences the learning process of skull anatomy on a similar level as traditional tools such as textbooks or plastic skull models.
Increasing the frequency of learning instances and interactions with models and specimens are advantages of teaching anatomy in AR\cite{Romand2020}.
Kolla et al.\cite{kolla2020medical} examined the effectiveness of VR technology in anatomy education and compared it to traditional teaching methods like lectures and cadaveric dissection.
28 first-year medical students used a VR headset to identify anatomical structures, drew them on a virtual skeleton, and then provided feedback through surveys. 
Their results indicated that VR was highly supported by the students, demonstrating its potential as a valuable tool for learning human anatomy and as a useful complement.
However, all study conditions were conducted within VR settings.
In a different study, the AR controlled group that was randomly selected from a biochemistry course suggests that AR educational apps motivated them to understand the visualized processes\cite{barrow_augmented_2019}.
Despite the popularity of using VR/AR methods in anatomy education, a review paper \cite{chytas2022extended} of 152 articles did not identify conclusive evidence of their efficiency over traditional anatomy education methods.


\subsection{Measures of Anatomy Learning Experience}
\noindent To analyze the learning experience and evaluate the effectiveness of VR/AR applications, various methods for data collection and different measures were used in previous research.
Kurniawan and Witjaksono~\cite{kurniawan2018human} evaluated the usefulness of the mobile-based AR application by employing the attitude questionnaire to analyze user perception. 
Nainggolan et al.~\cite{nainggolan2020user} evaluated the interactivity level of VR controller based on the user's agreement and satisfaction levels, and found that the use of the VR controller in the anatomy learning system was very interactive and satisfactory.
Tanjung et al.~\cite{fahmi2020comparison} conducted a comparison learning experience study to evaluate the level of acceptability and satisfaction towards three anatomical learning systems.
In our previous AR anatomy learning research ~\cite{barmaki_enhancement_2019,bork_empirical_2017,barmaki2020deep, bork2017exploring} and another study by Duncan‐Vaidya and Stevenson~\cite{duncan2021effectiveness} pre- and post-knowledge quizzes, and a usability questionnaires were used for data collection and analysis of the effectiveness of the AR tool.




\subsection{Gender Effects in Education}

\noindent In different education domains, gender differences have been discussed in recent studies.
Understanding gender effects in education is not conclusive, and it varies based on different disciplines and tasks \cite{fernandez-sanz_analysis_2012,wegge_age_2008}.
In the science, technology, engineering, and mathematics (STEM) domain, some non-positive learning experiences for females have been reported due to gender biases at the technical level~\cite{meadows_interactive_2015,nagappan_improving_2003}.
In the business domain, females' higher management ability in group tasks was highlighted \cite{bear_role_2011,de_paola_teamwork_2018,eagly_female_2003}.
Previous research has also shown that during cognitive tests, females have better information-processing skills than males \cite{rabbitt_unique_1995,schaie_age_1993}.
In Andersson's study \cite{andersson_net_2001}
on explicit spatial and verbal collaborative memory performance, the author reported better retention performance for females, but he argued that there was no main gender effect on team performance. 
Prinsen et al.\cite{prinsen_gender-related_2007}
noted that, in computer-mediated communication \cite{herring_computer-mediated_1996}
and computer-supported collaborative learning settings \cite{lehtinen_computer_1999,lipponen_challenges_1999,barmaki2020deep}, 
learning performance for different genders might change by various role distributions. 
In this paper, we will discuss the ethnographic observations related to gender differences (from individual and group composition standpoints. We need to acknowledge that we recruited students from a large laboratory classroom with pre-assigned teams thus, it was not feasible to form ethnographically balanced teams for all study conditions.

\section{Methods and Materials}\label{Sec:Method}

\noindent In this section, we will describe the details of the conducted study with our proposed tablet-based 3D visualization and screen-based AR learning tools.
Relevant hypotheses were established and evaluated to address our general research questions introduced in Section~\ref{Sec:Intro}.
This study was approved by the Institutional Review Board (Protocol \#HIRB00005021). 

\subsection{Digital Anatomy Learning Tools}
\label{Sec:ToolsAndIntervention}

\begin{figure}[tb]
\centering
\subfloat[]{\includegraphics[width=0.16\textwidth]{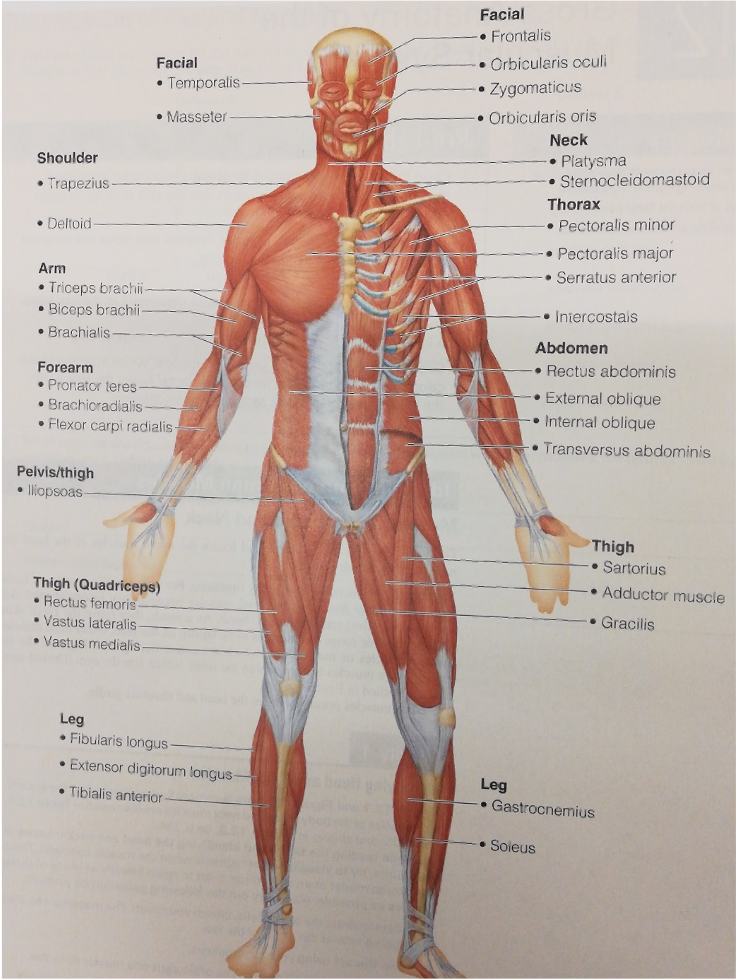}} 
\subfloat[]{\includegraphics[width=0.16\textwidth]{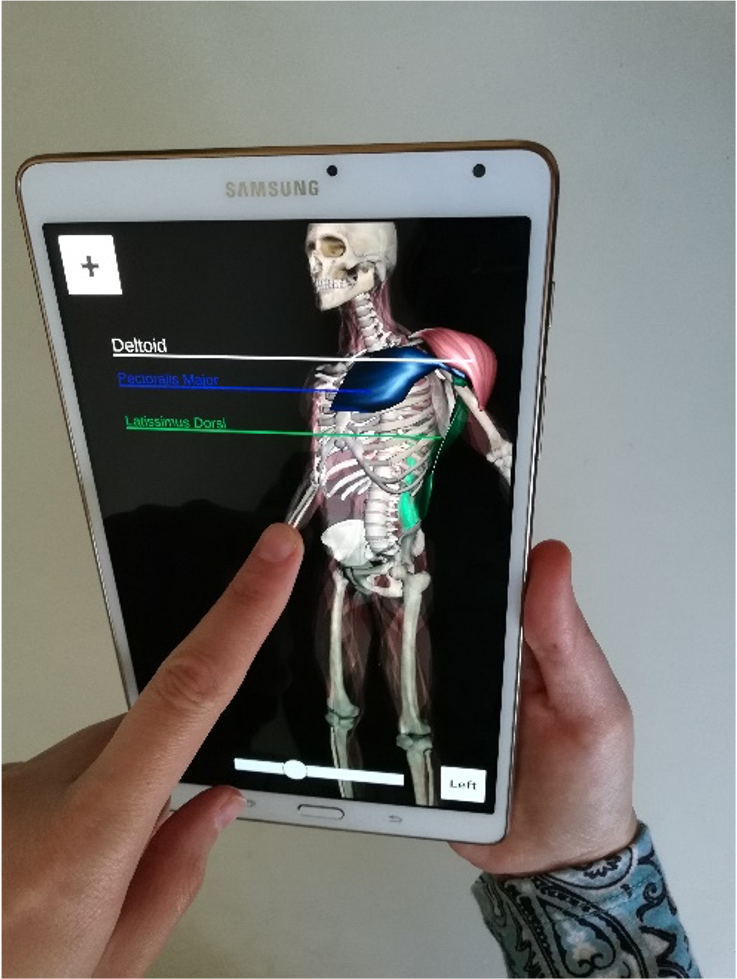}}
\subfloat[]{\includegraphics[width=0.16\textwidth]{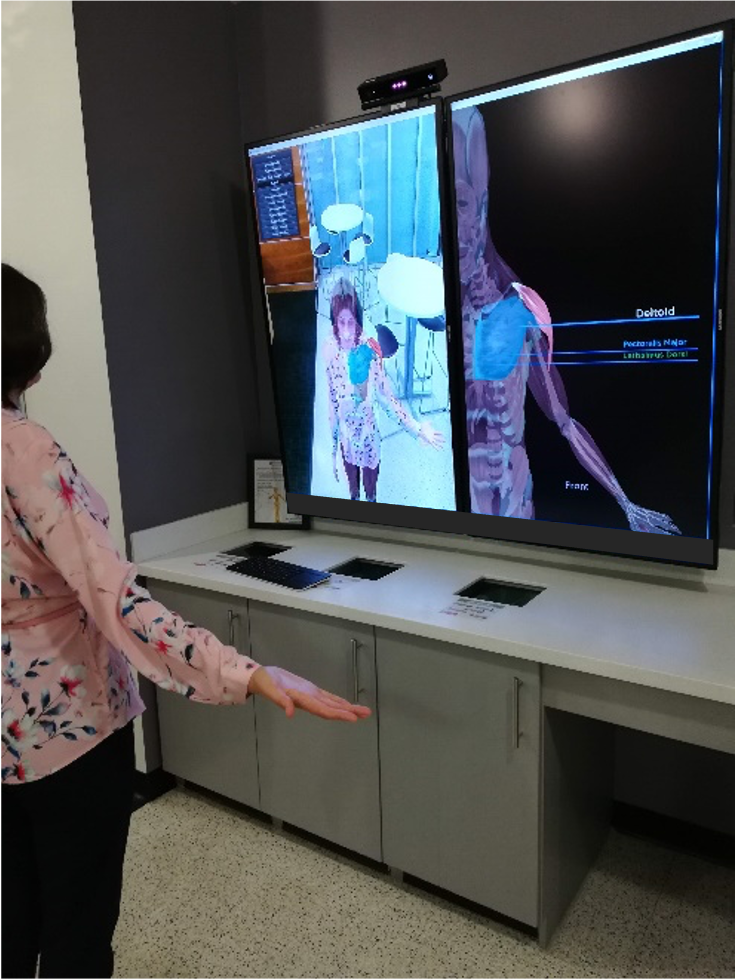}}\\
\subfloat[]{\includegraphics[width=0.16\textwidth]{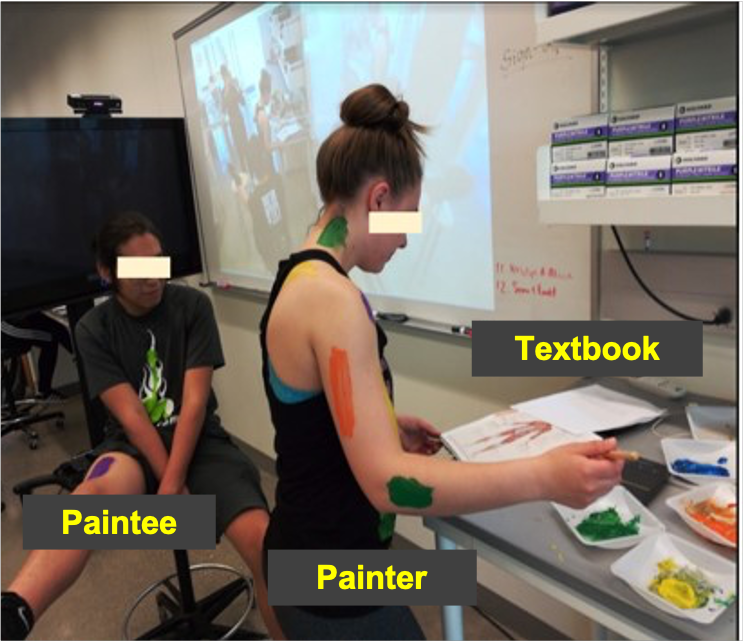}} 
\subfloat[]{\includegraphics[width=0.16\textwidth]{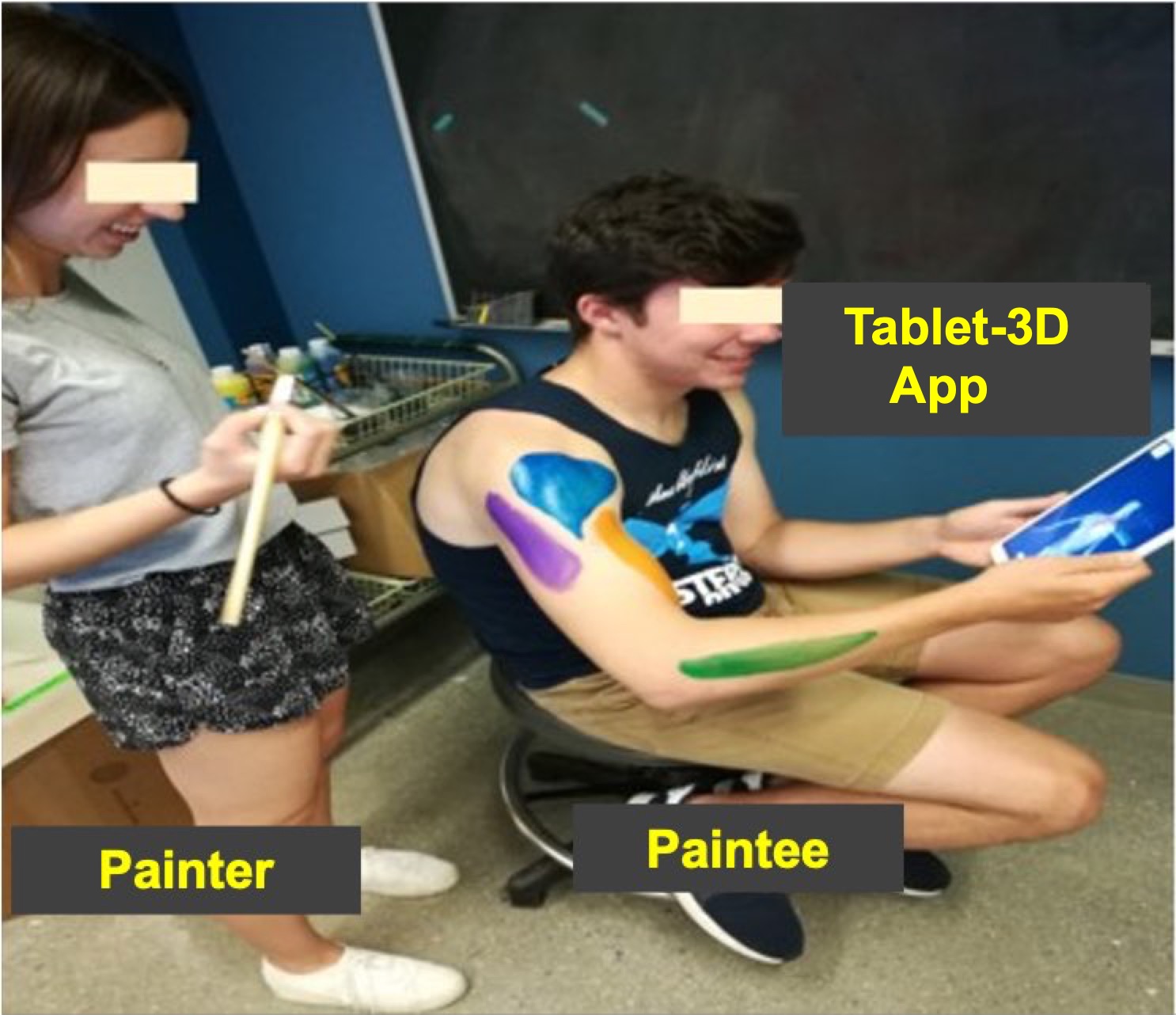}}
\subfloat[]{\includegraphics[width=0.16\textwidth]{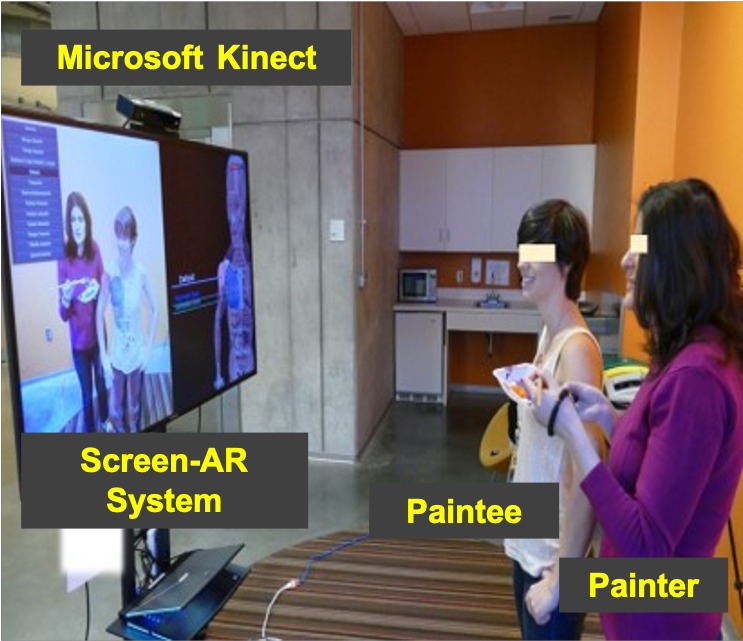}}\\
\caption{Three study conditions with different learning tools: (a) Textbook, (b) Tablet-3D, and (c) Screen-AR, and the images that captured the actual anatomy learning interventions with each of those learning tools: muscle painting activity in (d) Textbook, (e) Tablet-3D, and (f) Screen-AR.}
\label{fig:tools}
\vspace{-1em}
\end{figure}

\noindent To investigate the effects of tablet- and AR-based learning tools in anatomy education, we prepared in-house tablet-based 3D visualization and screen-based AR for our user study. 
The learning tools are presented in Figure~\ref{fig:tools}.
For Tablet-3D, we developed an interactive Android application for hand-held devices, which visualizes 3D virtual models of body muscles and labels on a tablet (see Figure~\ref{fig:tools}(b)).
The visualization of the 3D anatomy included regional and full-body anatomy models.
Participants could rotate, adjust zoom levels, toggle between color-coded virtual labels for enhancing readability, and highlight relevant muscle groups for our user study (see Figure~\ref{fig:tools}(c)).
The developed application was installed on Samsung Galaxy Tab 2 with a 10.1-inch screen.

For Screen-AR, we deployed our screen-based AR tool onto a large TV display with a mounted Kinect v2 camera. 
Screen-AR used a split-screen to show an AR view on the left side and a focused view of the anatomical model on the right side. 
Furthermore, body tracking allowed Screen-AR to superimpose 3D life-size virtual models of human anatomical systems, such as the muscular system, and labels atop the participant's mirrored body.
Participants could navigate between the relevant muscle groups using a hand-held clicker.
Depending on the muscle group, the virtual camera responsible for rendering the scene on the right screen automatically follows and zooms into the selected muscles.
Screen-AR consists of an Alienware 17 Laptop with GTX 1070m GPU, a 55-inch Samsung TV, and a Microsoft Kinect v2 tracking sensor, and all devices were mounted on a mobile TV cart.
We predefined the same muscle groups for Tablet-3D and Screen-AR for consistency.

\subsection{Team-based Learning Intervention}

\noindent After consultation with our laboratory instructor, Dr. Pearlman about the possibilities of testing our anatomy learning tools with her large classroom, we were introduced to a mandatory team-based lab activity about human muscle painting\cite{marieb_essentials_2006}.
Muscle painting, as a form of body painting, has been shown to be one of the common exercises in a medical curriculum\cite{mcmenamin_body_2008, barmaki_enhancement_2019}, and students execute the muscle painting activity typically with printed 2D visualization of human musculature from the lab manual. So, we prepared our learning tools based on these muscle painting activities, aiming to use them as a replacement for the 2D anatomy textbook and evaluate their learning efficacy and student performance.
During the intervention, participants worked in a team of two people to collaboratively learn and teach among peers about human muscle groups through a body painting activity.
They were asked to identify and paint major muscle parts on their body with washable painting supplies while using one of a randomly assigned (see Section \ref{Sec:StudyDesign}) learning tool.
Once one of the participants completed the painter's role, they switched their role to be a paintee with the learning partners.
We showed 40 muscles and labels to the participants to provide extensive anatomical landmarks. 
However, we only asked them to identify 12 major muscles (same for all conditions) and use appropriate body paint to colorize the muscles' location on their arms and legs.





\subsection{Study Design}
\label{Sec:StudyDesign}

\noindent We conducted a user study using a between-subjects design with three \textit{learning tool} conditions.
A brief description of each \textit{learning tool }is explained in the following.
\begin{itemize}[leftmargin=*]
    \item \textbf{Textbook}: As a traditional learning method, participants used a textbook during the team-based anatomy learning intervention described in Section~\ref{Sec:ToolsAndIntervention}.
    They identified the corresponding muscle parts on their partner's body while checking the location of target muscles from the textbook (Figure~\ref{fig:tools}(a, d)).
    The textbook could be carried while performing the muscle painting activity.
    \item \textbf{Tablet-3D}: Participants used our tablet-based 3D anatomy application during the learning intervention.
    The participants could carry the tablet that visualizes 3D virtual anatomy models and labels, while performing the painting activity (Figure~\ref{fig:tools}(b, e)).
    \item \textbf{Screen-AR}: Participants used our large screen-based AR anatomy tool during the activity.
    They could see the virtual 3D models and labels directly overlaid on their own bodies through the large screen (Figure~\ref{fig:tools}(c, f)).
\end{itemize}
The study was performed alongside 17 anatomy lab sessions with a maximum capacity of 20 students in each lab.
Considering the large scale of the study, participants could be randomly assigned to one of the three learning tool conditions by scheduling a single condition for each lab session.
This assignment method is termed hierarchical or clustered randomization, which is a common practice in educational studies and clinical trials\cite{davis_application_2002}.

\subsection{Participants}

\noindent We initially recruited 319 student participants (male: 128, female: 191), although we needed to exclude a subset of the data (described in the following). The recruitment took place via an online flyer from a laboratory course of General Biology as part of undergraduate premedical curricula. The recruitment was part of a general biology lab course, and the students were randomly assigned by their instructor into teams of two to four to perform their assignments throughout the semester.
The initial anatomy learning intervention  had teams of sizes two to four, however, teams of size two were only considered for this work. Larger teams had different task distribution in team, thus excluded. So, our final subset data of interest included \textit{\textbf{236}} participants (95 males and 141 females; age $M\!=\!19.77$, $SD\!=\!1.81$) in \textbf{118} teams of size two 
within three age-balanced groups with the following  gender compositions. Male pairs: 62 with age $M\!=\!19.45$, $SD\!=\!1.035$, Female pairs: 108 with age $M\!=\!19.71$, $SD\!=\!1.583$, and Mixed pairs: 66 with age $M\!=\!20.17$, $SD\!=\!2.527$). 
Table~\ref{Tab:Participants} shows the number of participants per condition.

\begin{table}[tb]
\caption{The number of participants in each study condition.}
\label{Tab:Participants}
\centering
\begin{tabular}{c|c|c|c}
\toprule
\textbf{Learning Tool} & \textbf{Male} & \textbf{Female} & \textbf{Total} \\
\midrule
Textbook  & 27   & 45     & 72    \\
Tablet-3D & 43   & 45     & 88    \\
Screen-AR & 25   & 51     & 76    \\
\midrule
\textbf{Total}     & 95   & 141    & 236  \\
\bottomrule
\end{tabular}
\vspace{-2ex}
\end{table}

\subsection{Procedure}



\noindent The study procedure was as follows. 
An online flyer was sent to all undergraduate students in the General Biology lab, inviting them to participate. Students needed to complete the body painting activity regardless of our study for their lab credits, but they could opt in to participate in our study to perform it slightly differently.
After the oral consent process, participants individually completed an online pre-questionnaire, which asked for their demographic information and evaluated their prior anatomy knowledge. 
Then they entered the learning intervention room with their pre-assigned teammates.
Each team only used one of the learning tools described in Section~\ref{Sec:StudyDesign} to complete the learning task, because of between-subjects study arrangements. 
During the painting activity, the participant in the painter's role tried to find appropriate muscles on the teammate's body (who played the role of a paintee) to be painted while navigating different muscle parts.
Participants in the Tablet-3D or textbook conditions used either the interactive application's 3D visualizations or the laboratory manual anatomy figures to complete the activity.
Particularly for the AR system, the participant in the paintee's role was asked to stand closer to the Kinect body tracking sensor to have appropriate digital anatomical illustrations superimposed on their body. 
The painter used the muscle overlay information on the TV screen to paint.
After completing the painting activity, all participants completed the online post-questionnaire individually, which asked about their interaction with the learning tools and evaluated their anatomy knowledge retention.
Before concluding the study session, they presented their painted limbs to their lab assistants.
The activity, including learning intervention and questionnaire completion, took approximately 20--30 minutes.

\subsection{Measures and Hypotheses}
\label{Sec:MeasuresHypotheses}


\noindent This section describes the measures we used for the study, which were collected through questionnaires.
We also introduce several hypotheses that we established based on the measures and our research questions in Section~\ref{Sec:Intro}.
We used online questionnaires on the Qualtrics platform (Provoto, UT) for designing and collecting pre- and post-questionnaires.


\subsubsection{Learning Experience}

We prepared nine subjective questions to examine the participants in the anatomy learning intervention using different learning tools.
The questions were presented on a five-point Likert scale (1: Strongly Disagree to 5: Strongly Agree) except for the ``willingness to recommend'' measure, which had a scale of 0 to 10 (10 means a strong willingness).
Our questions for each measure is described below. \textit{Textbook, Tablet-3D, or Screen-AR} was replaced by \textit{learning tool} statement, depending on the assigned condition.
\begin{itemize}[leftmargin=*]
    \item [-] \textbf{Easy to Paint}: ``Using \textit{learning tool} was easy and straightforward for completing the muscle painting activity.''
    \item [-] \textbf{Easy to Find Muscles}: ``I recognized and found the location of muscles easily using \textit{learning tool}.''
    \item [-] \textbf{Satisfaction}: ``Using \textit{learning tool} for the muscle painting activity was satisfying.''
    \item [-] \textbf{Learning Perception}: ``I learned a lot about muscle anatomy using/interacting with \textit{learning tool}.''
    \item [-] \textbf{Enjoyment}: ``I found using \textit{learning tool} to be enjoyable.''
    \item [-] \textbf{Effort to Focus}: ``I had to make an effort to keep my mind on the activity.''
    \item [-] \textbf{Lost Track of Time}: ``Time seemed to pass very quickly during the painting activity.''
    \item [-] \textbf{Willingness to Recommend}: ``I would recommend that my friends use \textit{learning tool} to study human muscles.''
    \item [-] \textbf{Learning Motivation}: ``Using \textit{learning tool} increased my enthusiasm for learning more about human anatomy.''
\end{itemize}

\noindent Considering the potential benefits of interactive visualizations in our 3D visualization tool, and more intuitive and direct information display on the real body in the AR tool \cite{blum2012mirracle}, we established the following hypotheses for the  measures:

\begin{itemize} [leftmargin=*]
    \item \textbf{H1}: The participants in the Tablet-3D condition or the Screen-AR condition will have more positive ratings than those in the Textbook condition for all the measures. \\\textit{\textbf{(Textbook $<$ Tablet-3D, Screen-AR)}}
    \item \textbf{H2}: The participants in the Screen-AR condition will further have more positive experience than those in the Tablet-3D condition for all the measures. \textit{\textbf{(Tablet-3D $<$ Screen-AR)}}
\end{itemize}



\noindent As Figure~\ref{fig:tools} shows, the anatomy models in all learning tools are male-based. 
Compared to female students, male students can easily recognize and access the muscles according to the same anatomical structures.
Also, given the prior research has shown that females tend to be more conservative about being touched~\cite{finn2010qualitative},
we established the following hypotheses to understand the role of the gender, gender composition:

\begin{itemize}[leftmargin=*]
\item \textbf{H3}: The male participants (male pairs) in the study will report more positive ratings than the female participants (females or mixed pairs), particularly in the measures like easy to paint and easy to find muscles.
\\\textit{\textbf{(Female $<$ Male)}} and \textit{\textbf{(Mixed $<=$ Females $<$ Males)}}
\end{itemize}

\subsubsection{Short-term Knowledge Retention}
To examine the learning performance in anatomy education, we evaluated the participant's anatomy knowledge retention.
We collected the participants' anatomy knowledge scores from pre- and post-tests, and calculated the score gain by subtracting the pre-test score from the post-test score.

\begin{itemize}[leftmargin=*]    
    \item [-] \textbf{Pre-Score}: The pre-test was a matching test with five questions in a provided diagram of the human anatomy muscle system. 
    In the anatomy diagram, 15 regions of the body were highlighted, and participants were asked to match five muscle labels provided in the pre-test with one of these 15 body regions.
    The number of correct matches was reported as the pre-test score in the range of [0, 5].
    \item [-] \textbf{Post-Score}: The post-test had a different (lateral) view of human anatomy, with a matching test similar to the pre-test.
    The number of correct matches was reported as the post-test score in the range of [0, 5].
    Both pre- and post-tests were designed and evaluated by anatomy instructors to be at the same level of difficulty.
    \item [-] \textbf{Score Gain}: Score gain, as the difference between pre- and post-test scores, was calculated in [-5, 5] range.
\end{itemize}

\noindent Based on the positive outcome that we anticipate in the Tablet-3D and Screen-AR conditions, we established the following hypotheses similar to H1 and H2 regarding the learning performance (or the improvement of anatomy knowledge retention):
\begin{itemize}[leftmargin=*]
    \item \textbf{H4}: The participants in the Tablet-3D condition or the Screen-AR condition will have a higher score gain than those in the Textbook condition. 
    \\\textit{\textbf{(Textbook $<$ Tablet-3D, Screen-AR)}}
    \item \textbf{H5}: The participants in the Screen-AR condition will further have a higher score gain than those in the Tablet-3D condition.   \textit{\textbf{  (Tablet-3D $<$ Screen-AR)}}
\end{itemize}

\section{Results}\label{Sec:Results}

\label{Sec:Results}

\noindent This section reports our analysis results considering the hypotheses we established in Section~\ref{Sec:MeasuresHypotheses}.
Since we are also interested in the possible effects of participant's gender on our measures, we have two factors to consider in our analysis: (1) learning tool and (2) participant's gender.
We first conducted two-way ANOVAs with these two factors to see if there is any interaction effect between the factors on both and performance measures.
We did not find any significant interactions between the learning tool and the participant gender; thus, we focused on the main effects of each factor using one-way ANOVAs ($\alpha\!=\!0.05$).
Multiple comparisons with Bonferroni correction were conducted for post-hoc tests.

\subsection{Learning Experience}
\label{Sec:UX_Results}

\noindent Here, we report the results of our analysis on the measures for both learning tool, participant gender, and group gender composition factors.
The detailed results for each measure are described below, and the overviews of the learning tool effects and the gender effects are summarized in Table~\ref{tab:my-table2}.

\unnumberedparagraph{\textbf{Easy to Paint}} The one-way ANOVA for the measure of ``easy to paint'' showed a significant effect of the learning tool ($F(2, 233)\!=\!7.786$, \textbf{\itshape{p}\,$=$\,0.0005}; $\eta_{\text{p}}^{2}\!=\!0.063$ - medium to large effect size). 
The post-hoc tests revealed that Tablet-3D ($M\!=\!4.67$, $SD\!=\!0.656$) had a higher score than Textbook ($M\!=\!4.26$, $SD\!=\!0.872$; \textbf{\itshape{p}\,$=$\,0.009}) or Screen-AR ($M\!=\!4.18$, $SD\!=\!1.016$; \textbf{\itshape{p}\,$=$\,0.001}).
This suggests that the participants felt the Tablet-3D condition was easier to perform the muscle painting activity than the other two conditions.
The analysis for the participant gender did not show any significant effect. 

\unnumberedparagraph{\textbf{Easy to Find Muscles}} For the ``easy to find muscles'' we did not find any significant effect of the learning tool.  
We found that this measure was significantly different among three group gender compositions ($F(2, 233)\!=\!5.62$, \textbf{\itshape{p}\,$=$\,0.004}; male pairs: $M\!=\!4.645$, $SD\!=\!0.630$; female pairs: $M\!=\!4.398$, $SD\!=\!0.875$; mixed pairs: $M\!=\!4.136$, $SD\!=\!1.006$).
Based on the post-hoc comparison with the Bonferroni test, the group with same gender compositions (males pairs and females pairs) were easier to perform the muscle painting activity than the mixed pairs (\textbf{\itshape{p}\,$=$\,0.005}).
Moreover, we found a significant effect of the participant's gender on this measure, too ($F(1, 234)\!=\!7.065$, \textbf{\itshape{p}\,$=$\,0.008}; $\eta_{\text{p}}^{2}\!=\!0.029$ - small to medium effect size).
The result showed that the male participants ($M\!=\!4.55$, $SD\!=\!0.632$) reported a significantly higher score for the easiness of finding muscles than the female participants ($M\!=\!4.26$, $SD\!=\!0.937$).

\begin{table}[tb]
\caption{The summary of Learning Experience inferential results. (* p\,$<$\,0.05, ** p\,$<$\,0.01, *** p\,$<$\,0.001.)}
\label{tab:my-table2}
\resizebox{\columnwidth}{!}{%
\begin{tabular}{lcccc}
\toprule
\multicolumn{1}{c}{\textbf{Variable}} & \textbf{F} & \textbf{p} & \textbf{Sig} & \textbf{$\eta_{p}^{2}$} \\ \midrule
Easy to Paint &  &  &  &  \\
\hspace*{0.2cm} \textit{Learning tool} & 7.786 & 0.0005 & *** & 0.063 \\
\hspace*{0.2cm} \textit{Gender} & 1.854 & 0.175 &  &  \\
Easy to Find Muscles &  &  &  &  \\
\hspace*{0.2cm} \textit{Gender} & 7.065 & 0.008 & ** & 0.029 \\
\hspace*{0.2cm} \textit{Gender composition} & 5.62 & 0.004 & ** &  \\
Satisfaction &  &  &  &  \\
\hspace*{0.2cm} \textit{Learning tool} & 0.207 & 0.813 &  &  \\
\hspace*{0.2cm} \textit{Gender} & 1.564 & 0.212 &  &  \\
\hspace*{0.2cm} \textit{Gender composition} & 3.91 & 0.021 & * &  \\
Learning Perception &  &  &  &  \\
\hspace*{0.2cm} \textit{Learning tool} & 0.009 & 0.991 &  &  \\
\hspace*{0.2cm} \textit{Gender} & 6.245 & 0.013 & * & 0.026 \\
\hspace*{0.2cm} \textit{Gender composition} & 3.63 & 0.028 & * &  \\
Enjoyment &  &  &  &  \\
\hspace*{0.2cm} \textit{Learning tool} & 9.87 & 0.0001 & *** & 0.078 \\
\hspace*{0.2cm} \textit{Gender} & 10.008 & 0.002 & ** & 0.041 \\
\hspace*{0.2cm} \textit{Gender composition} & 4.03 & 0.019 & * &  \\
Effort to Focus &  &  &  &  \\
\hspace*{0.2cm} \textit{Learning tool} & 1.679 & 0.189 &  &  \\
\hspace*{0.2cm} \textit{Gender} & 0.598 & 0.44 &  &  \\
\hspace*{0.2cm} \textit{Gender composition} & 1.84 & 0.161 &  &  \\
Lost Track of Time &  &  &  &  \\
\hspace*{0.2cm} \textit{Learning tool} & 0.332 & 0.718 &  &  \\
\hspace*{0.2cm} \textit{Gender} & 0.413 & 0.521 &  &  \\
\hspace*{0.2cm} \textit{Gender composition} & 0.36 & 0.696 &  &  \\
Willingness to Recommend &  &  &  &  \\
\hspace*{0.2cm} \textit{Learning tool} & 4.596 & 0.011 & * & 0.038 \\
\hspace*{0.2cm} \textit{Gender} & 4.323 & 0.039 & * & 0.018 \\
\hspace*{0.2cm} \textit{Gender composition} & 1.19 & 0.306 &  &  \\
Learning Motivation &  &  &  &  \\
\hspace*{0.2cm} \textit{Learning tool} & 7.441 & 0.0007 & *** & 0.06 \\
\hspace*{0.2cm} \textit{Gender} & 4.907 & 0.028 & * & 0.021 \\
\hspace*{0.2cm} \textit{Gender composition} & 2.95 & 0.055 &  &  \\ \bottomrule
\end{tabular}%
}
\vspace{-2em}
\end{table}
    
\unnumberedparagraph{\textbf{Satisfaction}} We did not find any significant effects of both the learning tool and the participant gender. 
However, this ``satisfaction'' measure was significantly different in gender compositions ($F(2, 233)\!=\!3.91$, \textbf{\itshape{p}\,$=$\,0.021}). 
Among three different gender compositions (male pairs: $M\!=\!4.484$, $SD\!=\!0.671$; female pairs: $M\!=\!4.259$, $SD\!=\!0.921$; mixed pairs: $M\!=\!4.045$, $SD\!=\!0.999$), Bonferroni test showed that male pairs had significantly higher satisfaction than mixed pairs (\textbf{\itshape{p}\,$=$\,0.017}).
    
\unnumberedparagraph{\textbf{Learning Perception}} For this 
measure, we did not find a significant effect of the learning tool, 
but there was a main effect of the participant gender with a statistical significance ($F(1, 234)\!=\!6.245$, \textbf{\itshape{p}\,$=$\,0.013}; $\eta_{\text{p}}^{2}\!=\!0.026$ - small to medium effect size).
    This showed again that the male participants ($M\!=\!4.08$, $SD\!=\!0.834$) had a higher score than the female participants ($M\!=\!3.76$, $SD\!=\!1.062$).
Also, a significant effect for the ``learning perception'' measure was observed ($F(2, 233)\!=\!3.63$, \textbf{\itshape{p}\,$=$\,0.028}) among gender compositions of male pairs ($M\!=\!4.177$, $SD\!=\!0.758$), female pairs ($M\!=\!3.778$, $SD\!=\!1.026$), and mixed pairs ($M\!=\!3.803$, $SD\!=\!1.084$).
Bonferroni test indicated that male pairs had a significantly higher score than female pairs (\textbf{\itshape{p}\,$=$\,0.034}).

\unnumberedparagraph{\textbf{Enjoyment}} For the ``enjoyment'' we found main effects of the learning tool, the participant gender, and group gender composition factors.
For the learning tool ($F(2, 233)\!=\!9.870$, \textbf{\itshape{p}\,$=$\,0.0001}; $\eta_{\text{p}}^{2}\!=\!0.078$ - medium to large effect size), we further compared the conditions and found that Textbook ($M\!=\!3.78$, $SD\!=\!1.051$) had a lower score than Tablet-3D ($M\!=\!4.32$, $SD\!=\!0.865$; \textbf{\itshape{p}\,$=$\,0.001}) and Screen-AR ($M\!=\!4.41$, $SD\!=\!0.897$; \textbf{\itshape{p}\,$=$\,0.001}).
For the gender effect ($F(1, 234)\!=\!10.008$, \textbf{\itshape{p}\,$=$\,0.002}; $\eta_{\text{p}}^{2}\!=\!0.041$ - small to medium effect size), the result showed that the male participants ($M\!=\!4.42$, $SD\!=\!0.807$) had a higher score than the female participants ($M\!=\!4.02$, $SD\!=\!1.038$).
For the gender composition (male pairs: $M\!=\!4.468$, $SD\!=\!0.804$; female pairs: $M\!=\!4.037$, $SD\!=\!1.049$; mixed pairs: $M\!=\!4.152$, $SD\!=\!0.932$), a significant difference was observed ($F(2, 233)\!=\!4.03$, \textbf{\itshape{p}\,$=$\,0.019}).
The Bonferroni post-hoc test showed that the male pairs had a significantly higher score than the female pairs (\textbf{\itshape{p}\,$=$\,0.016}).

\unnumberedparagraph{\textbf{Effort to Focus}} There were no significant effects found in the ``effort to focus'' measure.

\unnumberedparagraph{\textbf{Lost Track of Time}} For learning tool and gender-related factors, there were also no significant effects for this measure.

\unnumberedparagraph{\textbf{Willingness to Recommend}} We found a main effect of the learning tool for the ``willingness to recommend'' ($F(2, 233)\!=\!4.596$, \textbf{\itshape{p}\,$=$\,0.011}; $\eta_{\text{p}}^{2}\!=\!0.038$ - small to medium effect size).
The post-hoc tests revealed that Textbook ($M\!=\!6.96$, $SD\!=\!2.185$) had a lower score than Tablet-3D ($M\!=\!7.82$, $SD\!=\!2.003$; \textbf{\itshape{p}\,$=$\,0.024}) and Screen-AR ($M\!=\!7.83$, $SD\!=\!1.865$; \textbf{\itshape{p}\,$=$\,0.028}).
The gender effect was also found for this ($F(1, 234)\!=\!4.323$, \textbf{\itshape{p}\,$=$\,0.039}; $\eta_{\text{p}}^{2}\!=\!0.018$ - small to medium effect size), which showed that the male participants ($M\!=\!7.89$, $SD\!=\!1.825$) had a higher score than the female participants ($M\!=\!7.33$, $SD\!=\!2.164$).
On the contrary, no group gender composition effect was identified.

\unnumberedparagraph{\textbf{Learning Motivation}} 
We found the main effects of the learning tool and participant gender.
The effect of the learning tool was significant ($F(2, 233)\!=\!7.441$, \textbf{\itshape{p}\,$=$\,0.0007}; $\eta_{\text{p}}^{2}\!=\!0.060$ - medium effect size), and the post-hoc tests showed that Textbook ($M\!=\!3.57$, $SD\!=\!1.032$) had a lower score than Tablet-3D ($M\!=\!3.94$, $SD\!=\!0.987$; \textbf{\itshape{p}\,$=$\,0.044}) and Screen-AR ($M\!=\!4.17$, $SD\!=\!0.839$; \textbf{\itshape{p}\,$=$\,0.001}).
The gender effect was also significant ($F(1, 234)\!=\!4.907$, \textbf{\itshape{p}\,$=$\,0.028}; $\eta_{\text{p}}^{2}\!=\!0.021$ - small to medium effect size), showing that the male participants ($M\!=\!4.07$, $SD\!=\!0.902$) had a higher score than the female participants ($M\!=\!3.79$, $SD\!=\!1.020$).
Based on Bonferroni test, we further compared the pairs and found that the male pairs had a significantly higher score than the female pairs for ``learning motivation'' (\textbf{\itshape{p}\,$=$\,0.048}). 
    



\subsection{Learning Performance}

\label{Sec:Performance_Results}

To ensure valid knowledge retention evaluation, we analyzed pre-score differences among learning tool conditions and gender groups. This involved assessing the balance in prior anatomy knowledge within learning tool groups and gender compositions. We did not find any differences in the pre-score measure among the learning tool conditions. However, we found a significant difference between the male and female participants (\textbf{\itshape{p}\,$=$\,0.019}), which showed the male participants ($M\!=\!2.05$, $SD\!=\!1.283$) had a higher pre-score than the female participants ($M\!=\!1.69$, $SD\!=\!1.070$).
While controlling the pre-score as a covariate, we conducted one-way ANCOVA to investigate the effects of the learning tool, the participant's gender, or the group gender composition; however, no significant results were found.
Among different learning tool conditions, compared with the score gain ($M\!=\!0.069$, $SD\!=\!1.437$) and the post-score ($M\!=\!1.92$, $SD\!=\!1.441$) in the Textbook condition, participants in the Tablet-3D condition achieved the highest score gain ($M\!=\!0.10$, $SD\!=\!1.447$) with post-score ($M\!=\!1.82$, $SD\!=\!1.282$); Screen-AR groups achieved the second highest score gain ($M\!=\!0.079$, $SD\!=\!1.512$) with post-score ($M\!=\!2.04$, $SD\!=\!1.501$). 
With respect to the participant's gender, females achieved a higher score gain ($M\!=\!0.20$, $SD\!=\!1.455$) with post-score ($M\!=\!1.89$, $SD\!=\!1.342$) than males' score gain ($M\!=\!-0.084$, $SD\!=\!1.456$) with post-score ($M\!=\!1.97$, $SD\!=\!1.491$).
Among three group gender compositions, compared with the score gain ($M\!=\!-0.145$, $SD\!=\!1.401$) and the post-score ($M\!=\!1.887$, $SD\!=\!1.307$) of the male pairs, female groups achieved the highest score gain ($M\!=\!0.269$, $SD\!=\!1.464$) with post-score ($M\!=\!1.852$, $SD\!=\!1.288$); mixed pairs achieved the second highest score gain ($M\!=\!0$, $SD\!=\!1.488$) with post-score ($M\!=\!2.06$, $SD\!=\!1.654$). 
There are no main or interaction effects on the post-score and the score gain among the three factors: learning tool, gender, and group gender composition.

\section{Discussion}

\label{Sec:Discussion}


\noindent In this section, we summarize our findings based on the results while connecting them to our established hypotheses. We also discuss the justifications and implications of our findings and associate them with previous research.
\unnumberedparagraph{\textbf{Digital Tools Improved Learning Experience}} We found several significant effects of the learning tool for some of our measures, which we reported in Section~\ref{Sec:UX_Results}.
In general, the results show that the Tablet-3D and the Screen-AR could provide more positive experience scores than the Textbook, e.g., for the measures of ``enjoyment,'' ``willingness to recommend,'' and ``learning motivation,'' which partly supports our \textbf{H1}.
These results align with previous literature findings that showed 3D visualization technologies increased students' engagement in anatomy learning\cite{hackett_effect_2018,luursema_role_2008}.
The positive scores in some of our  measures could be related to the novelty of the VR/AR technology as an anatomy learning tool, and the intuitive and interactive 3D models could also be an important factor in influencing the perception of the learning experience.
Some qualitative comments from the participants in the Tablet-3D condition or the Screen-AR condition post-session also support our reasoning for the positive scores. For example, some of the participants in Tablet-3D or Screen-AR said:
\begin{quote}
\leftskip=-10pt
\rightskip=-10pt
\textit{``Interface could be refined a little; the model was great.''}\\
\textit{``It was an enjoyable and educational experience.''}\\
\textit{``Super cool to be the model and see my body's muscles.''}\\
\textit{``It was fun to paint a muscle and helped put it into the context of reality instead of simply seeing it on a page.''}
\end{quote}

\noindent Interestingly, however, we found no significant benefits of the Screen-AR compared to the Tablet-3D, which we expected in our \textbf{H2}.
Instead, we found a higher score in the Tablet-3D condition than the Screen-AR condition for the ``easy to paint'' measure.
Based on the participants' feedback below, we realized that some participants complained about the intermittent misalignment of virtual contents in the Screen-AR condition.

\begin{quote}
\leftskip=-10pt
\rightskip=-10pt
\textit{``The calibration was not great and it was hard to actually see the muscle projected onto myself and my partner.''}\\
\textit{``Sometimes it was hard to see which muscle was being pointed.''}\\
\textit{``It was a little hard to distinguish where in the muscle was sometimes because the overlay wasn't actually exact/not proportional to my body.''}
\end{quote}

\noindent This implies that the participants were quite sensitive to the accuracy and reliability of body tracking for AR content registration, which could be crucial for the learning experience.
Some major contributors to this AR accuracy issue include the 3D virtual models, which followed a male anatomy limb proportions rather than female, and the physical proximity of the painter and paintee in front of the body tracking system.  


\unnumberedparagraph{\textbf{No Influence on Learning Performance}} Positive effects of our digital tools on learning performance based on prior research was expected~\cite{maresky_virtual_2019,nicholson_can_2006, barmaki_enhancement_2019}.
As we reported in Section\ref{Sec:Performance_Results}, there were no significant effects of the learning tool or the participant's gender for the learning performance, e.g., the anatomy knowledge retention, which means no evidence to support our \textbf{H4} and \textbf{H5}.
However, prior research showed that positive experience could increase learning performance and retention~\cite{maresky_virtual_2019,nicholson_can_2006}, and our proposed 3D visualization and AR learning tools did promote more positive compared to the traditional textbook-based learning.
In that sense, we are inspired to conduct further research on different learning analytic metrics.
Additionally, the intervention that the participants had in our study was only a one-time session for about 10--15 minutes, which could not be enough to reveal the positive effects of AR on learning performance.
As most learning science studies require longitudinal studies for their efficacy, we will propose a longitudinal study in future.


\unnumberedparagraph{\textbf{Male Users Noted More Positive Experience}} Beyond investigating the learning tool effects, we also established an interesting question about gender effects in anatomy education, as introduced in \textbf{H3} in Section~\ref{Sec:MeasuresHypotheses}.
We found significant effects of the participant gender on various measures: ``easy to find muscles,'' ``learning perception,'' ``enjoyment,'' ``willingness to recommend,'' and ``learning motivation.''
For all those measures, male participants reported higher (more positive) scores than female participants, which partly supports our \textbf{H3}.
 Considering gender composition, similar patterns were observed on the ``easy to find muscles,'' ``learning perception,'' and ``enjoyment'' measures. 
 When comparing gender compositions, we found statistically significant differences between male pairs and mixed pairs, as well as between male pairs and female pairs. Male pairs reported the highest learning experience scores. These finding also provides partial support for \textbf{H3}. We should note that this effect was not associated with a specific \textit{learning tool}.
There could be some aspects of our learning intervention that female participants did not like compared to male participants, which seemed to be the \textit{painting part}, not the \textit{learning tool}. Female participants in our study might be more conservative about the body painting activity than males. 
According to prior research~\cite{cookson2018exploration,finn2010qualitative}, unequal engagement of students in body painting activities may be due to cultural, social, and religious barriers concerning body image, nudity, gender, vulnerability, and embarrassment.
Not surprisingly, such tendencies were reflected in the after-session feedback from some female participants.

\begin{quote}
\leftskip=-10pt
\rightskip=-10pt
\textit{``Painting was unnecessary, but it was cool to see the muscles on my body.''}
\textit{``[The activity may] take into account clothes and girls.''}\\
\textit{``I am not a fan of body painting. However, I love the program and think that it is really cool.'''}\\
\textit{``I don't like having paint on my skin.''}
\end{quote}

\noindent These observations emphasize the importance of gender perspectives in the design. It also recommends exploring a new learning module, one that doesn't involve body painting.


\unnumberedparagraph{\textbf{Future Work and Limitations}}
The overarching goal of our study was to explore and understand the possible effects of using advanced learning technologies on anatomy learning and student performance. Identifying anatomical parts in a real body, as in our current study, is primarily acknowledged in clinical anatomy. While not immediately applicable to sessions involving cadavers or anatomy models, exploring this avenue, such as using a Kinect sensor positioned above a patient bed, presents an interesting future path. Our discussion unveils that we require multi-session, long-term, and modified studies to draw more conclusive findings for knowledge retention and task completion time. As noted, some students liked the anatomy learning tools but disliked the body painting activity, so the selection of body painting as baseline activity may have been a confounding factor. In addition, this paper covered dyadic teams and we plan to analyze our data with participants in larger teams while controlling for demographics of age, gender, and prior knowledge level to form more uniform teams. We also plan to expand student learning analysis by collecting and using multi-modal measurements for more effective and objective \textit{learning tool} evaluations.

\section{Conclusion}
\label{Sec:Conclusions}

\noindent This paper investigated the effects of digital anatomy learning tools, using our in-house tablet-based 3D visualization system and the screen-based AR application overlaying 3D anatomy structures onto the digital mirror image of students. 
We conducted a large-scale study with \textbf{236} premedical students, and compared our digital learning tools with conventional textbook.
Our results indicated that both digital learning tools could improve the learning experience, particularly in ``enjoyment'', ``willingness to recommend'', and ``learning motivation''. 
We also found that male participants generally reported more positive experiences than females.
Our findings include further considerations of learners' gender and performance in body painting anatomy education.
Based on the enhanced, progressive short-term learning performance and enjoyment in our study, we expect a similar trend in the long-term retention of future 3D technology.



\section*{Acknowledgment}
\noindent We wish to express our gratitude to the study participants and laboratory assistants. We wish to acknowledge the support from the support of the Johns Hopkins University Science of
Learning Institute, and National Science Foundation for award \#2321274. 
Any opinions, findings, and conclusions expressed in this material are those of the authors.

\balance
\bibliographystyle{IEEEtran}
\bibliography{IEEEabrv,bibfile}

\begin{thebibliography}{10}
\providecommand{\url}[1]{#1}
\csname url@samestyle\endcsname
\providecommand{\newblock}{\relax}
\providecommand{\bibinfo}[2]{#2}
\providecommand{\BIBentrySTDinterwordspacing}{\spaceskip=0pt\relax}
\providecommand{\BIBentryALTinterwordstretchfactor}{4}
\providecommand{\BIBentryALTinterwordspacing}{\spaceskip=\fontdimen2\font plus
\BIBentryALTinterwordstretchfactor\fontdimen3\font minus \fontdimen4\font\relax}
\providecommand{\BIBforeignlanguage}[2]{{%
\expandafter\ifx\csname l@#1\endcsname\relax
\typeout{** WARNING: IEEEtran.bst: No hyphenation pattern has been}%
\typeout{** loaded for the language `#1'. Using the pattern for}%
\typeout{** the default language instead.}%
\else
\language=\csname l@#1\endcsname
\fi
#2}}
\providecommand{\BIBdecl}{\relax}
\BIBdecl

\bibitem{BLANCHARD200573}
S.~Blanchard, ``Anatomy and physiology,'' in \emph{Introduction to Biomedical Engineering (Second Edition)}, second edition~ed., ser. Biomedical Engineering, J.~D. Enderle, S.~M. Blanchard, and J.~D. Bronzino, Eds.\hskip 1em plus 0.5em minus 0.4em\relax Boston: Academic Press, 2005, pp. 73--125.

\bibitem{o2008development}
P.~J. O’Byrne, A.~Patry, and J.~A. Carnegie, ``The development of interactive online learning tools for the study of anatomy,'' \emph{Medical Teacher}, vol.~30, no.~8, pp. e260--e271, 2008.

\bibitem{green2018relationship}
R.~A. Green, L.~Y. Whitburn, A.~Zacharias, G.~Byrne, and D.~L. Hughes, ``The relationship between student engagement with online content and achievement in a blended learning anatomy course,'' \emph{Anatomical sciences education}, vol.~11, no.~5, pp. 471--477, 2018.

\bibitem{chan2019approaches}
A.~Y.-C.~C. Chan, O.~T. Cate, E.~J. Custers, M.~S. van Leeuwen, and R.~L. Bleys, ``Approaches of anatomy teaching for seriously resource-deprived countries: A literature review.'' \emph{Education for Health: Change in Learning \& Practice}, vol.~32, no.~2, 2019.

\bibitem{leung2020modernising}
B.~C. Leung, M.~Williams, C.~Horton, and T.~D. Cosker, ``Modernising anatomy teaching: Which resources do students rely on?'' \emph{Journal of medical education and curricular development}, vol.~7, pp. 1--7, 2020.

\bibitem{leung_anatomy_2006}
K.-k. Leung, K.-S. Lu, T.-S. Huang, and B.-S. Hsieh, ``Anatomy instruction in medical schools: connecting the past and the future,'' \emph{Advances in Health Sciences Education}, vol.~11, no.~2, pp. 209--215, 2006, publisher: Springer.

\bibitem{winkelmann_anatomical_2007}
A.~Winkelmann, ``Anatomical dissection as a teaching method in medical school: a review of the evidence,'' \emph{Medical education}, vol.~41, no.~1, pp. 15--22, 2007, publisher: Wiley Online Library.

\bibitem{lovis_mixed_2020}
C.~Lovis, ``Mixed and {Augmented} {Reality} {Tools} in the {Medical} {Anatomy} {Curriculum},'' \emph{Digital Personalized Health and Medicine: Proceedings of MIE 2020}, vol. 270, pp. 322--326, 2020, publisher: IOS Press.

\bibitem{marmulla_augmented_2005}
R.~Marmulla, H.~Hoppe, J.~Mühling, and G.~Eggers, ``An augmented reality system for image-guided surgery: {This} article is derived from a previous article published in the journal {International} {Congress} {Series},'' \emph{International journal of oral and maxillofacial surgery}, vol.~34, no.~6, pp. 594--596, 2005, publisher: Elsevier.

\bibitem{Romand2020}
M.~Romand, D.~Dugas, C.~Gaudet-Blavignac, J.~Rochat, and C.~Lovis, ``{Mixed and augmented reality tools in the medical anatomy curriculum},'' \emph{Studies in Health Technology and Informatics}, vol. 270, no. June, pp. 322--326, 2020.

\bibitem{huk_who_2006}
T.~Huk, ``Who benefits from learning with {3D} models? {The} case of spatial ability,'' \emph{Journal of computer assisted learning}, vol.~22, no.~6, pp. 392--404, 2006, publisher: Wiley Online Library.

\bibitem{lipponen_challenges_1999}
L.~Lipponen, ``The challenges for computer supported collaborative learning in elementary and secondary level: {Finnish} perspectives,'' 1999, publisher: International Society of the Learning Sciences (ISLS).

\bibitem{stieff_mental_2007}
M.~Stieff, ``Mental rotation and diagrammatic reasoning in science,'' \emph{Learning and instruction}, vol.~17, no.~2, pp. 219--234, 2007, publisher: Elsevier.

\bibitem{cookson2018exploration}
N.~E. Cookson, J.~J. Aka, and G.~M. Finn, ``An exploration of anatomists’ views toward the use of body painting in anatomical and medical education: An international study,'' \emph{Anatomical sciences education}, vol.~11, no.~2, pp. 146--154, 2018.

\bibitem{Diaz2021learning}
C.~M. Diaz and T.~Woolley, ````learning by doing'': a mixed-methods study to identify why body painting can be a powerful approach for teaching surface anatomy to health science students,'' \emph{Medical Science Educator}, vol.~31, no.~6, pp. 1875--1887, 2021.

\bibitem{barmaki_enhancement_2019}
R.~Barmaki, K.~Yu, R.~Pearlman, R.~Shingles, F.~Bork, G.~M. Osgood, and N.~Navab, ``Enhancement of {Anatomical} {Education} {Using} {Augmented} {Reality}: {An} {Empirical} {Study} of {Body} {Painting},'' \emph{Anatomical sciences education}, vol.~12, no.~6, pp. 599--609, 2019.

\bibitem{nainggolan2020user}
F.~Nainggolan, B.~Siregar, and F.~Fahmi, ``User experience in using vive controller as a controller in anatomy learning system in virtual reality environment,'' in \emph{Journal of Physics: Conference Series}, vol. 1566, no.~1.\hskip 1em plus 0.5em minus 0.4em\relax IOP Publishing, 2020, p. 012096.

\bibitem{snelling_attitudes_2003}
J.~Snelling, A.~Sahai, and H.~Ellis, ``Attitudes of medical and dental students to dissection,'' \emph{Clinical Anatomy: The Official Journal of the American Association of Clinical Anatomists and the British Association of Clinical Anatomists}, vol.~16, no.~2, pp. 165--172, 2003, publisher: Wiley Online Library.

\bibitem{gunderman_exploring_2005}
R.~B. Gunderman and P.~K. Wilson, ``Exploring the human interior: {The} roles of cadaver dissection and radiologic imaging in teaching anatomy,'' \emph{Academic Medicine}, vol.~80, no.~8, pp. 745--749, 2005, publisher: LWW.

\bibitem{trelease_going_2000}
R.~B. Trelease, G.~L. Nieder, J.~Dørup, and M.~S. Hansen, ``Going virtual with {QuickTime} {VR}: new methods and standardized tools for interactive dynamic visualization of anatomical structures,'' \emph{The Anatomical Record: An Official Publication of the American Association of Anatomists}, vol. 261, no.~2, pp. 64--77, 2000, publisher: Wiley Online Library.

\bibitem{mclachlan_teaching_2004}
J.~C. McLachlan, J.~Bligh, P.~Bradley, and J.~Searle, ``Teaching anatomy without cadavers,'' \emph{Medical education}, vol.~38, no.~4, pp. 418--424, 2004, publisher: Wiley Online Library.

\bibitem{dehoff_learning_2011}
M.~E. DeHoff, K.~L. Clark, and K.~Meganathan, ``Learning outcomes and student-perceived value of clay modeling and cat dissection in undergraduate human anatomy and physiology,'' \emph{Advances in physiology education}, vol.~35, no.~1, pp. 68--75, 2011, publisher: American Physiological Society Bethesda, MD.

\bibitem{hays_spatial_1996}
T.~A. Hays, ``Spatial abilities and the effects of computer animation on short-term and long-term comprehension,'' \emph{Journal of educational computing research}, vol.~14, no.~2, pp. 139--155, 1996, publisher: SAGE Publications Sage CA: Los Angeles, CA.

\bibitem{chickness2022novel}
J.~P. Chickness, K.~M. Trautman-Buckley, K.~Evey, and L.~Labranche, ``Novel development of a 3d digital mediastinum model for anatomy education,'' \emph{Translational Research in Anatomy}, vol.~26, p. 100158, 2022.

\bibitem{yammine_meta-analysis_2015}
K.~Yammine and C.~Violato, ``A meta-analysis of the educational effectiveness of three-dimensional visualization technologies in teaching anatomy,'' \emph{Anatomical sciences education}, vol.~8, no.~6, pp. 525--538, 2015, publisher: Wiley Online Library.

\bibitem{golenhofen_use_2019}
N.~Golenhofen, F.~Heindl, C.~Grab-Kroll, D.~A. Messerer, T.~M. Böckers, and A.~Böckers, ``The use of a mobile learning tool by medical students in undergraduate anatomy and its effects on assessment outcomes,'' \emph{Anatomical sciences education}, 2019, publisher: Wiley Online Library.

\bibitem{lemos_design_2019}
R.~R. Lemos, C.~M. Rudolph, A.~V. Batista, K.~R. Conceição, P.~F. Pereira, B.~S. Bueno, P.~J. Fiuza, and S.~S. Mansur, ``Design of a {Web3D} {Serious} {Game} for {Human} {Anatomy} {Education}: {A} {Web3D} {Game} for {Human} {Anatomy} {Education},'' in \emph{Handbook of {Research} on {Immersive} {Digital} {Games} in {Educational} {Environments}}.\hskip 1em plus 0.5em minus 0.4em\relax IGI Global, 2019, pp. 586--611.

\bibitem{maresky_virtual_2019}
H.~Maresky, A.~Oikonomou, I.~Ali, N.~Ditkofsky, M.~Pakkal, and B.~Ballyk, ``Virtual reality and cardiac anatomy: {Exploring} immersive three-dimensional cardiac imaging, a pilot study in undergraduate medical anatomy education,'' \emph{Clinical Anatomy}, vol.~32, no.~2, pp. 238--243, 2019, publisher: Wiley Online Library.

\bibitem{donnelly2009virtual}
L.~Donnelly, D.~Patten, P.~White, and G.~Finn, ``Virtual human dissector as a learning tool for studying cross-sectional anatomy,'' \emph{Medical teacher}, vol.~31, no.~6, pp. 553--555, 2009.

\bibitem{keenan2019integrating}
I.~D. Keenan and A.~ben Awadh, ``Integrating 3d visualisation technologies in undergraduate anatomy education,'' \emph{Biomedical Visualisation: Volume 1}, pp. 39--53, 2019.

\bibitem{barrack2015step}
D.~Barrack, D.~Horn, and B.~Benninger, ``A step by step visual guide to using the sectra visualization table for 1st and 2nd year medical students,'' \emph{The FASEB Journal}, vol.~29, pp. 692--8, 2015.

\bibitem{lim_use_2016}
K.~H.~A. Lim, Z.~Y. Loo, S.~J. Goldie, J.~W. Adams, and P.~G. McMenamin, ``Use of {3D} printed models in medical education: a randomized control trial comparing {3D} prints versus cadaveric materials for learning external cardiac anatomy,'' \emph{Anatomical sciences education}, vol.~9, no.~3, pp. 213--221, 2016, publisher: Wiley Online Library.

\bibitem{bork_empirical_2017}
F.~Bork, R.~Barmaki, U.~Eck, K.~Yu, C.~Sandor, and N.~Navab, ``Empirical study of non-reversing magic mirrors for augmented reality anatomy learning,'' in \emph{2017 {IEEE} {International} {Symposium} on {Mixed} and {Augmented} {Reality} ({ISMAR})}.\hskip 1em plus 0.5em minus 0.4em\relax IEEE, 2017, pp. 169--176.

\bibitem{silva_emerging_2018}
J.~N. Silva, M.~Southworth, C.~Raptis, and J.~Silva, ``Emerging applications of virtual reality in cardiovascular medicine,'' \emph{JACC: Basic to Translational Science}, vol.~3, no.~3, pp. 420--430, 2018, number: 3 Publisher: JACC: Basic to Translational Science.

\bibitem{bacca2014augmented}
J.~L. Bacca~Acosta, S.~M. Baldiris~Navarro, R.~Fabregat~Gesa, S.~Graf \emph{et~al.}, ``Augmented reality trends in education: a systematic review of research and applications,'' \emph{Journal of Educational Technology and Society, 2014, vol. 17, n{\'u}m. 4, p. 133-149}, 2014.

\bibitem{hackett_effect_2018}
M.~Hackett and M.~Proctor, ``The effect of autostereoscopic holograms on anatomical knowledge: a randomised trial,'' \emph{Medical education}, vol.~52, no.~11, pp. 1147--1155, 2018, number: 11 Publisher: Wiley Online Library.

\bibitem{jacob2012lindsay}
C.~Jacob, S.~von Mammen, T.~Davison, A.~Sarraf-Shirazi, V.~Sarpe, A.~Esmaeili, D.~Phillips, I.~Yazdanbod, S.~Novakowski, S.~Steil \emph{et~al.}, ``Lindsay virtual human: Multi-scale, agent-based, and interactive,'' in \emph{Advances in Intelligent Modelling and Simulation}.\hskip 1em plus 0.5em minus 0.4em\relax Springer, 2012, pp. 327--349.

\bibitem{luursema_role_2008}
J.-M. Luursema, W.~B. Verwey, P.~A. Kommers, and J.-H. Annema, ``The role of stereopsis in virtual anatomical learning,'' \emph{Interacting with Computers}, vol.~20, no. 4-5, pp. 455--460, 2008, publisher: Oxford University Press Oxford, UK.

\bibitem{preim2018survey}
B.~Preim and P.~Saalfeld, ``A survey of virtual human anatomy education systems,'' \emph{Computers \& Graphics}, vol.~71, pp. 132--153, 2018.

\bibitem{duncan-vaidya_effectiveness_2020}
E.~Duncan-Vaidya and E.~Stevenson, ``The {Effectiveness} of an {Augmented} {Reality} {Head}-{Mounted} {Display} in {Learning} {Skull} {Anatomy} at a {Community} {College}.'' \emph{Anatomical Sciences Education}, 2020.

\bibitem{kolla2020medical}
S.~Kolla, M.~Elgawly, J.~P. Gaughan, and E.~Goldman, ``Medical student perception of a virtual reality training module for anatomy education,'' \emph{Medical Science Educator}, vol.~30, pp. 1201--1210, 2020.

\bibitem{barrow_augmented_2019}
J.~Barrow, C.~Forker, A.~Sands, D.~O'Hare, and W.~Hurst, ``Augmented {Reality} for {Enhancing} {Life} {Science} {Education},'' in \emph{{VISUAL} 2019-{The} {Fourth} {International} {Conference} on {Applications} and {Systems} of {Visual} {Paradigms}}, 2019.

\bibitem{chytas2022extended}
D.~Chytas, M.~Piagkou, T.~Demesticha, G.~Tsakotos, and K.~Natsis, ``Are extended reality technologies (erts) more effective than traditional anatomy education methods?'' \emph{Surgical and Radiologic Anatomy}, vol.~44, no.~9, pp. 1215--1218, 2022.

\bibitem{kurniawan2018human}
M.~H. Kurniawan, G.~Witjaksono \emph{et~al.}, ``Human anatomy learning systems using augmented reality on mobile application,'' \emph{Procedia Computer Science}, vol. 135, pp. 80--88, 2018.

\bibitem{fahmi2020comparison}
F.~Fahmi, K.~Tanjung, F.~Nainggolan, B.~Siregar, N.~Mubarakah, and M.~Zarlis, ``Comparison study of user experience between virtual reality controllers, leap motion controllers, and senso glove for anatomy learning systems in a virtual reality environment,'' in \emph{IOP Conference Series: Materials Science and Engineering}, vol. 851, no.~1.\hskip 1em plus 0.5em minus 0.4em\relax IOP Publishing, 2020, p. 012024.

\bibitem{barmaki2020deep}
R.~Barmaki and Z.~Guo, ``Deep neural networks for collaborative learning analytics: Evaluating team collaborations using student gaze point prediction,'' \emph{Australasian Journal of Educational Technology}, vol.~36, no.~6, pp. 53--71, 2020.

\bibitem{bork2017exploring}
F.~Bork, R.~Barmaki, U.~Eck, P.~Fallavolita, B.~Fuerst, and N.~Navab, ``Exploring non-reversing magic mirrors for screen-based augmented reality systems,'' in \emph{2017 IEEE virtual reality (VR)}.\hskip 1em plus 0.5em minus 0.4em\relax IEEE, 2017, pp. 373--374.

\bibitem{duncan2021effectiveness}
E.~A. Duncan-Vaidya and E.~L. Stevenson, ``The effectiveness of an augmented reality head-mounted display in learning skull anatomy at a community college,'' \emph{Anatomical Sciences Education}, vol.~14, no.~2, pp. 221--231, 2021.

\bibitem{fernandez-sanz_analysis_2012}
L.~Fernandez-Sanz and S.~Misra, ``Analysis of cultural and gender influences on teamwork performance for software requirements analysis in multinational environments,'' \emph{IET software}, vol.~6, no.~3, pp. 167--175, 2012, publisher: IET.

\bibitem{wegge_age_2008}
J.~Wegge, C.~Roth, B.~Neubach, K.-H. Schmidt, and R.~Kanfer, ``Age and gender diversity as determinants of performance and health in a public organization: the role of task complexity and group size.'' \emph{Journal of Applied Psychology}, vol.~93, no.~6, p. 1301, 2008, publisher: American Psychological Association.

\bibitem{meadows_interactive_2015}
L.~A. Meadows, D.~Sekaquaptewa, and M.~C. Paretti, ``Interactive panel: {Improving} the experiences of marginalized students on engineering design teams,'' in \emph{{ASEE} {Annual} {Conference} \& {Exposition}: {Excellence} in {Education}}, vol.~26, 2015, p.~1.

\bibitem{nagappan_improving_2003}
N.~Nagappan, L.~Williams, M.~Ferzli, E.~Wiebe, K.~Yang, C.~Miller, and S.~Balik, ``Improving the {CS1} experience with pair programming,'' \emph{ACM SIGCSE Bulletin}, vol.~35, no.~1, pp. 359--362, 2003, publisher: ACM New York, NY, USA.

\bibitem{bear_role_2011}
J.~B. Bear and A.~W. Woolley, ``The role of gender in team collaboration and performance,'' \emph{Interdisciplinary science reviews}, vol.~36, no.~2, pp. 146--153, 2011, publisher: Taylor \& Francis.

\bibitem{de_paola_teamwork_2018}
M.~De~Paola, F.~Gioia, and V.~Scoppa, ``Teamwork, {Leadership} and {Gender},'' IZA Discussion Papers, Tech. Rep., 2018.

\bibitem{eagly_female_2003}
A.~H. Eagly and L.~L. Carli, ``The female leadership advantage: {An} evaluation of the evidence,'' \emph{The leadership quarterly}, vol.~14, no.~6, pp. 807--834, 2003, number: 6 Publisher: Elsevier.

\bibitem{rabbitt_unique_1995}
P.~Rabbitt, C.~Donlan, P.~Watson, L.~McInnes, and N.~Bent, ``Unique and interactive effects of depression, age, socioeconomic advantage, and gender on cognitive performance of normal healthy older people.'' \emph{Psychology and aging}, vol.~10, no.~3, p. 307, 1995, number: 3 Publisher: American Psychological Association.

\bibitem{schaie_age_1993}
K.~W. Schaie and S.~L. Willis, ``Age difference patterns of psychometric intelligence in adulthood: generalizability within and across ability domains.'' \emph{Psychology and aging}, vol.~8, no.~1, p.~44, 1993, publisher: American Psychological Association.

\bibitem{andersson_net_2001}
J.~Andersson, ``Net effect of memory collaboration: {How} is collaboration affected by factors such as friendship, gender and age?'' \emph{Scandinavian journal of Psychology}, vol.~42, no.~4, pp. 367--375, 2001, number: 4 Publisher: Wiley Online Library.

\bibitem{prinsen_gender-related_2007}
F.~R. Prinsen, M.~L. Volman, and J.~Terwel, ``Gender-related differences in computer-mediated communication and computer-supported collaborative learning,'' \emph{Journal of Computer Assisted Learning}, vol.~23, no.~5, pp. 393--409, 2007, number: 5 Publisher: Wiley Online Library.

\bibitem{herring_computer-mediated_1996}
S.~C. Herring, \emph{Computer-mediated communication: {Linguistic}, social, and cross-cultural perspectives}.\hskip 1em plus 0.5em minus 0.4em\relax John Benjamins Publishing, 1996, vol.~39.

\bibitem{lehtinen_computer_1999}
E.~Lehtinen, K.~Hakkarainen, L.~Lipponen, M.~Rahikainen, and H.~Muukkonen, ``Computer supported collaborative learning: {A} review,'' \emph{The JHGI Giesbers reports on education}, vol.~10, p. 1999, 1999, publisher: University of Nijmegen Nijmegen.

\bibitem{marieb_essentials_2006}
E.~N. Marieb and P.~B. Jackson, \emph{Essentials of {Human} {Anatomy} \& {Physiology} {Laboratory} {Manual}}.\hskip 1em plus 0.5em minus 0.4em\relax Pearson/Benjamin Cummings, 2006.

\bibitem{mcmenamin_body_2008}
P.~G. McMenamin, ``Body painting as a tool in clinical anatomy teaching,'' \emph{Anatomical sciences education}, vol.~1, no.~4, pp. 139--144, 2008.

\bibitem{davis_application_2002}
L.~Davis, F.~G. Hamza-Lup, J.~Daly, Y.~Ha, S.~Frolich, C.~Meyer, G.~Martin, J.~Norfleet, K.-C. Lin, C.~Imielinska, and {others}, ``Application of augmented reality to visualizing anatomical airways,'' in \emph{Helmet-and {Head}-{Mounted} {Displays} {VII}}, vol. 4711.\hskip 1em plus 0.5em minus 0.4em\relax International Society for Optics and Photonics, 2002, pp. 400--405.

\bibitem{blum2012mirracle}
T.~Blum, V.~Kleeberger, C.~Bichlmeier, and N.~Navab, ``mirracle: An augmented reality magic mirror system for anatomy education,'' in \emph{2012 IEEE Virtual Reality Workshops (VRW)}.\hskip 1em plus 0.5em minus 0.4em\relax IEEE, 2012, pp. 115--116.

\bibitem{finn2010qualitative}
G.~M. Finn and J.~C. McLachlan, ``A qualitative study of student responses to body painting,'' \emph{Anatomical sciences education}, vol.~3, no.~1, pp. 33--38, 2010.

\bibitem{nicholson_can_2006}
D.~T. Nicholson, C.~Chalk, W.~R.~J. Funnell, and S.~J. Daniel, ``Can virtual reality improve anatomy education? {A} randomised controlled study of a computer-generated three-dimensional anatomical ear model,'' \emph{Medical education}, vol.~40, no.~11, pp. 1081--1087, 2006, publisher: Wiley Online Library.

\end{thebibliography}

\end{document}